# Reduced form of a Mueller matrix

José J. Gil[1] and Ignacio San José[2]

[1]*Universidad de Zaragoza. Pedro Cerbuna 12, 50009 Zaragoza Spain*
[2]*Instituto Aragonés de Estadística. Gobierno de Aragón. Bernardino Ramazzini 5, 50015 Zaragoza, Spain*
*Corresponding author: ppgil@unizar.es*

**Abstract**

Through a simple procedure based on the Lu-Chipman decomposition [S-Y. Lu and R. C. Chipman, J. Opt. Soc. Am A 13, 1106 (1996)] any depolarizing Mueller matrix can be transformed into a reduced form which accumulates the depolarization and polarizance properties into a set of six parameters. The simple structure of this reduced form provides straightforward ways for the general characterization of Mueller matrices as well as for the analysis of singular Mueller matrices.

## 1 Introduction

Polarimetry has demonstrated to be a powerful tool for the study and characterization of countless kinds of material samples and therefore its applications are constantly increasing. Linear interactions producing changes in the state of polarization of the electromagnetic probe are characterized by the Mueller matrix of the corresponding medium (under the given interaction conditions, like the spectral profile of the probing beam, the angles of incidence and observation, etc.). It is well known that the algebraic structure of Mueller matrices is quite intricate, and therefore any advance in their representation by means of equivalent systems composed of basic elements, like retarders, diattenuators and simplified forms of depolarizers, has a high number of potential applications in experimental polarimetry.

This work is devoted to the definition and study of two *reduced forms* of a generic depolarizing Mueller matrix **M**, which are obtained through the sequential application to **M** of the *generalized polar decomposition* (also called *Lu-Chipman decomposition* or *forward decomposition*) [1] and the singular value decomposition of the low-right 3x3 submatrix of **M**.

The new approach is proved to be useful for the analysis of some polarimetric properties and thus complements other well established and powerful approaches like the Lu-Chipman decomposition itself (in its forward and reverse forms [1-3]), the symmetric decomposition [4] and the arrow decomposition [5,6].

To simplify most of mathematical expressions involving Mueller matrices along this paper, let us bring up the partitioned block expression of a Mueller matrix [7],

$$\mathbf{M} = m_{00} \begin{pmatrix} 1 & \mathbf{D}^T \\ \mathbf{P} & \mathbf{m} \end{pmatrix}$$

$$\mathbf{D} \equiv \frac{1}{m_{00}}(m_{01}, m_{02}, m_{03})^T \qquad \mathbf{P} \equiv \frac{1}{m_{00}}(m_{10}, m_{20}, m_{30})^T \qquad \mathbf{m} \equiv \frac{1}{m_{00}} \begin{pmatrix} m_{11} & m_{12} & m_{13} \\ m_{21} & m_{22} & m_{23} \\ m_{31} & m_{32} & m_{33} \end{pmatrix} \quad (1)$$

where the **D** and **P** are respectively called the *diattenuation vector* and the *polarizance vector* of **M** [1]. The absolute values of these vectors are called *diattenuation*, $D \equiv |\mathbf{D}|$, and *polarizance*, $P \equiv |\mathbf{P}|$.

Both polarizance $P$ and diattenuation $D$ play complementary roles depending on the direction of the probing electromagnetic wave (forward or reverse) [8]; $D$ is both the diattenuation of **M** and the polarizance of the *reverse Mueller matrix* $\mathbf{M}^r \equiv \mathrm{diag}(1,1,-1,1)\mathbf{M}^T \mathrm{diag}(1,1,-1,1)$ [9,10] corresponding to the same interaction as **M** but interchanging the input and output directions, while $P$ is the diattenuation of $\mathbf{M}^r$ besides being the polarizance of **M**.

A medium satisfying $P=1$ and $D<1$ is called a *depolarizing polarizer*; a medium with $D=1$ and $P<1$ is called a *depolarizing analyzer* and a medium with $D=P=1$ is called a *polarizer* (also *nondepolarizing polarizer* or *polarizer-analyzer*) [1].

The *degree of polarimetric purity* [8] of **M** is given by the *depolarization index* [11]

$$P_\Delta = \sqrt{D^2 + P^2 + \|\mathbf{m}\|_2^2}\Big/\sqrt{3} \tag{2}$$

where $\|\mathbf{m}\|_2 \equiv \mathrm{tr}(\mathbf{m}^T\mathbf{m})$ stands for the Frobenius norm of the submatrix **m** of **M**. Further, the *average intensity coefficient* of **M** (i.e., transmittance or reflectance for unpolarized input states) is given by $m_{00}$.

Mueller matrices associated with systems that do not depolarize any totally polarized input state (i.e., whose depolarization index satisfies $P_\Delta = 1$) are called *pure Mueller matrices* (also *nondepolarizing* or *Jones-Mueller matrices*), while Mueller matrices satisfying $P_\Delta < 1$ are called *nonpure* or *depolarizing* Mueller matrices.

Next, we summarize other theoretical subjects that are necessary to address the definition and properties of the reduced forms of **M**.

The Lu-Chipman decomposition is formulated as [1]

$$\mathbf{M} \equiv m_{00}\begin{pmatrix} 1 & \mathbf{D}^T \\ \mathbf{P} & \mathbf{m} \end{pmatrix} = m_{00}\hat{\mathbf{M}}_{\Delta P}\mathbf{M}_R\hat{\mathbf{M}}_D;$$

$$\hat{\mathbf{M}}_{\Delta P} \equiv \begin{pmatrix} 1 & \mathbf{0}^T \\ \mathbf{P}_{\Delta P} & \mathbf{m}_{\Delta P} \end{pmatrix}, \quad \mathbf{M}_R \equiv \begin{pmatrix} 1 & \mathbf{0}^T \\ \mathbf{0} & \mathbf{m}_R \end{pmatrix}, \quad \hat{\mathbf{M}}_D \equiv \begin{pmatrix} 1 & \mathbf{D}^T \\ \mathbf{D} & \mathbf{m}_D \end{pmatrix}, \tag{3}$$

$$\mathbf{P}_{\Delta P} = \frac{\mathbf{P} - \mathbf{m}\mathbf{D}}{1-D^2}, \quad \mathbf{m}_R^{-1} = \mathbf{m}_R^T \ (\det\mathbf{m}_R = +1), \quad \mathbf{m}_D \equiv \sqrt{1-D^2}\,\mathbf{I}_3 + \left(1-\sqrt{1-D^2}\right)\hat{\mathbf{D}}\otimes\hat{\mathbf{D}}^T,$$

$$\mathbf{I}_3 \equiv \mathrm{diag}(1,1,1), \quad \hat{\mathbf{D}} \equiv \mathbf{D}/D,$$

where the normalized symmetric matrix $\hat{\mathbf{M}}_D$ represents a diattenuator whose diattenuation-polarizance vector **D** is equal to the diattenuation vector of **M**; the orthogonal matrix $\mathbf{M}_R$ represents a retarder, and the normalized depolarizer matrix $\hat{\mathbf{M}}_{\Delta P}$, with $\mathbf{m}_{\Delta P}^T = \mathbf{m}_{\Delta P}$, represents a depolarizer with nonzero polarizance and zero diattenuation. The product $\mathbf{M}_R\hat{\mathbf{M}}_D$ corresponds to a pure Mueller matrix, which can also be submitted to the reverse form of the polar decomposition [12] $\mathbf{M}_R\hat{\mathbf{M}}_D = \hat{\mathbf{M}}'_D\mathbf{M}_R$, with $\hat{\mathbf{M}}'_D = \mathbf{M}_R\hat{\mathbf{M}}_D\mathbf{M}_R^T$ (note that $\hat{\mathbf{M}}_D$ and $\hat{\mathbf{M}}'_D$ have equal polarizance-diattenuation D). For a pure Mueller matrix, the equality $\mathbf{P} = \mathbf{m}\mathbf{D}$ is satisfied and the depolarizer degenerates into the identity matrix.

In passing, it is worth to recall that the decomposition (3) can also be expressed as [2]

$$\mathbf{M} = m_{00} \, \hat{\mathbf{M}}_{\Delta P} \, \mathbf{M}_R \, \hat{\mathbf{M}}_D = m_{00} \, \hat{\mathbf{M}}_{\Delta P} \, \hat{\mathbf{M}}'_D \, \mathbf{M}_R = m_{00} \, \mathbf{M}_R \, \hat{\mathbf{M}}'_{\Delta P} \, \hat{\mathbf{M}}_D$$
$$\left( \hat{\mathbf{M}}'_D \equiv \mathbf{M}_R \, \hat{\mathbf{M}}_D \, \mathbf{M}_R^T \qquad \hat{\mathbf{M}}'_{\Delta P} \equiv \mathbf{M}_R^T \, \hat{\mathbf{M}}_{\Delta P} \, \mathbf{M}_R \right) \tag{4}$$

By applying the decomposition (3) to the Mueller matrix $\mathbf{M}^T$ (which, as indicated above is closely related to the reverse Mueller matrix $\mathbf{M}^r$), we obtain

$$\mathbf{M}^T \equiv m_{00} \begin{pmatrix} 1 & \mathbf{P}^T \\ \mathbf{D} & \mathbf{m}^T \end{pmatrix} = m_{00} \hat{\mathbf{M}}_{\Delta D}^T \, \mathbf{M}_R^T \, \hat{\mathbf{M}}_P \tag{5}$$

where $\hat{\mathbf{M}}_{\Delta D}^T$ is a depolarizer without diattenuation, $\mathbf{M}_R^T$ is the transposed matrix of $\mathbf{M}_R$ (i.e., $\mathbf{M}_R^T$ represents a retarder), and $\hat{\mathbf{M}}_P$ represents a diattenuator. Therefore, the following *reverse decomposition* of $\mathbf{M}$ is obtained by taking the transpose in Eq. (5) [3] (recall that $\hat{\mathbf{M}}_P = \hat{\mathbf{M}}_P^T$)

$$\mathbf{M} = m_{00} \, \hat{\mathbf{M}}_P \, \mathbf{M}_R \, \hat{\mathbf{M}}_{\Delta D}$$
$$\hat{\mathbf{M}}_{\Delta D} \equiv \begin{pmatrix} 1 & \mathbf{D}_{\Delta D}^T \\ \mathbf{0} & \mathbf{m}_{\Delta D} \end{pmatrix} \quad \mathbf{M}_R \equiv \begin{pmatrix} 1 & \mathbf{0}^T \\ \mathbf{0} & \mathbf{m}_R \end{pmatrix} \quad \hat{\mathbf{M}}_P \equiv \begin{pmatrix} 1 & \mathbf{P}^T \\ \mathbf{P} & \mathbf{m}_P \end{pmatrix} \tag{6}$$

where the normalized depolarizer $\hat{\mathbf{M}}_{\Delta D}$ exhibits nonzero diattenuation and zero polarizance, and the retarder $\mathbf{M}_R$ coincides with that of the forward decomposition.

The reverse decomposition (6) can also be transformed into the following alternative forms [3]

$$\mathbf{M} = m_{00} \, \hat{\mathbf{M}}_P \, \mathbf{M}_R \, \hat{\mathbf{M}}_{\Delta D} = m_{00} \, \mathbf{M}_R \, \hat{\mathbf{M}}'_P \, \hat{\mathbf{M}}_{\Delta D} = m_{00} \, \hat{\mathbf{M}}_P \, \hat{\mathbf{M}}'_{\Delta D} \, \mathbf{M}_R$$
$$\left( \hat{\mathbf{M}}'_P \equiv \mathbf{M}_R^T \, \hat{\mathbf{M}}_P \, \mathbf{M}_R \qquad \hat{\mathbf{M}}'_{\Delta D} \equiv \mathbf{M}_R \, \hat{\mathbf{M}}_{\Delta D} \, \mathbf{M}_R^T \right) \tag{7}$$

The matrix $\hat{\mathbf{M}}_D$ is fully determined by the diattenuation vector $\mathbf{D}$ of $\mathbf{M}$, whereas the calculation of $\hat{\mathbf{M}}_{\Delta P}$ and $\mathbf{M}_R$ requires to distinguish between nonsingular and singular Mueller matrices. The general procedure for the calculation of $\hat{\mathbf{M}}_{\Delta P}$, $\mathbf{M}_R$ and $\hat{\mathbf{M}}_D$, including the case of $\mathbf{M}$ being singular, can be found in Ref. [1]. Moreover, the calculation of $\hat{\mathbf{M}}_{\Delta D}$, $\mathbf{M}_R$ and $\hat{\mathbf{M}}_P$ is achieved by applying the same procedure to $\mathbf{M}^T$.

**2 Reduced forms of a Mueller matrix**

Since the symmetric matrix $\mathbf{m}_{\Delta P}$ can be diagonalized as $\mathbf{m}_{\Delta P} = \mathbf{m}_{Rf} \, \mathbf{m}_{L\Delta P} \, \mathbf{m}_{Rf}^T$, $\mathbf{m}_{Rf}$ being an orthogonal matrix, the forward decomposition (3), $\mathbf{M} = m_{00} \, \hat{\mathbf{M}}_{\Delta P} \, \mathbf{M}_R \, \hat{\mathbf{M}}_D$, can be transformed in the following manner:

$$\mathbf{M} = m_{00} \left( \mathbf{M}_{Rf} \, \hat{\mathbf{M}}_{L\Delta P} \, \mathbf{M}_{Rf}^T \right) \mathbf{M}_R \, \hat{\mathbf{M}}_D = m_{00} \, \mathbf{M}_{Rf} \, \hat{\mathbf{M}}_{L\Delta P} \, \hat{\mathbf{M}}_{JD}$$
$$\hat{\mathbf{M}}_{L\Delta P} \equiv \begin{pmatrix} 1 & \mathbf{0}^T \\ \mathbf{P}_{L\Delta P} & \mathbf{m}_{L\Delta P} \end{pmatrix}, \quad \mathbf{M}_{Rf} \equiv \begin{pmatrix} 1 & \mathbf{0}^T \\ \mathbf{0} & \mathbf{m}_{Rf} \end{pmatrix}, \quad \hat{\mathbf{M}}_{JD} \equiv \mathbf{M}_{Rf}^T \, \mathbf{M}_R \, \hat{\mathbf{M}}_D,$$
$$\mathbf{m}_{L\Delta P} \equiv \mathrm{diag}(l_{\Delta P1}, l_{\Delta P2}, \varepsilon \, l_{\Delta P3}), \quad 0 \leq l_{\Delta P3} \leq l_{\Delta P2} \leq l_{\Delta P1}, \quad \varepsilon \equiv (\det \mathbf{M})/|\det \mathbf{M}|, \tag{8}$$
$$\mathbf{P}_{L\Delta P} = \mathbf{m}_{Rf} \, \mathbf{P}_{\Delta P} = \frac{1}{1-D^2} \, \mathbf{m}_{Rf}^T \left( \mathbf{P} - \mathbf{mD} \right),$$

where the only depolarizing component is $\hat{\mathbf{M}}_{L\Delta P}$, hereafter called the *forward reduced form* of $\mathbf{M}$, which has only seven nonzero elements located on its first column and on its diagonal:

$$\hat{\mathbf{M}}_{L\Delta P} = \begin{pmatrix} 1 & 0 & 0 & 0 \\ P_{L\Delta P1} & l_{\Delta P1} & 0 & 0 \\ P_{L\Delta P2} & 0 & l_{\Delta P2} & 0 \\ P_{L\Delta P3} & 0 & 0 & \varepsilon l_{\Delta P3} \end{pmatrix} \quad (9)$$

In analogy to the forward decomposition, the reverse decomposition can be expressed as follows

$$\mathbf{M} = m_{00}\, \hat{\mathbf{M}}_P\, \mathbf{M}_R \left( \mathbf{M}_{Rr}\, \hat{\mathbf{M}}_{L\Delta D}\, \mathbf{M}_{Rr}^T \right) = m_{00}\, \hat{\mathbf{M}}_{JP}\, \hat{\mathbf{M}}_{L\Delta D}\, \mathbf{M}_{Rr}^T,$$

$$\hat{\mathbf{M}}_{L\Delta D} \equiv \begin{pmatrix} 1 & \mathbf{D}_{L\Delta D}^T \\ \mathbf{0} & \mathbf{m}_{L\Delta D} \end{pmatrix}, \quad \mathbf{M}_{Rr} \equiv \begin{pmatrix} 1 & \mathbf{0}^T \\ \mathbf{0} & \mathbf{m}_{Rr} \end{pmatrix}, \quad \hat{\mathbf{M}}_{JP} \equiv \hat{\mathbf{M}}_P\, \mathbf{M}_R\, \mathbf{M}_{Rr},$$

$$\mathbf{m}_{L\Delta D} \equiv \operatorname{diag}(l_{\Delta D1}, l_{\Delta D2}, \varepsilon\, l_{\Delta D3}), \quad 0 \le l_{\Delta D3} \le l_{\Delta D2} \le l_{\Delta D1}, \quad \varepsilon \equiv (\det \mathbf{M})/|\det \mathbf{M}|,$$

$$\mathbf{D}_{L\Delta D} = \mathbf{m}_{Rr}\, \mathbf{D}_{\Delta D} = \frac{1}{1-P^2}\, \mathbf{m}_{Rr}\left(\mathbf{D} - \mathbf{m}^T \mathbf{P}\right),$$

(10)

where the only nonpure component is the depolarizer $\hat{\mathbf{M}}_{L\Delta D}$, hereafter called the *reverse reduced form* of **M**, whose only nonzero elements are located on its diagonal and on its first row:

$$\hat{\mathbf{M}}_{L\Delta D} = \begin{pmatrix} 1 & D_{L\Delta D1} & D_{L\Delta D2} & D_{L\Delta D3} \\ 0 & l_{\Delta D1} & 0 & 0 \\ 0 & 0 & l_{\Delta D2} & 0 \\ 0 & 0 & 0 & \varepsilon l_{\Delta D3} \end{pmatrix} \quad (11)$$

## 3 Eigenstates of the reduced Mueller matrices

The notion of eigenstates of a Mueller matrix **M** refers to the Stokes vectors **s** satisfying $\mathbf{M}\mathbf{s} = k\,\mathbf{s}$ with $k > 0$. Other eigenvectors of **M** that do not satisfy this condition are unphysical. In fact, due to the block structure of a Mueller matrix **M**, where the vectors

$$\mathbf{s}_{D+} \equiv m_{00}\begin{pmatrix} 1 \\ \mathbf{D} \end{pmatrix}, \quad \mathbf{s}_{P+} \equiv m_{00}\begin{pmatrix} 1 \\ \mathbf{P} \end{pmatrix}, \quad (12)$$

satisfy the conditions to be considered as Stokes vectors, **M** has at most two eigenstates, the remaining eigenvectors of **M** being unphysical.

Due to their particular structures, the reduced forms have the following respective unique eigenstates (eigenvectors with the form of Stokes vectors)

$$\mathbf{s}_{L\Delta P} = \begin{pmatrix} 1 \\ P_{L\Delta P1}/(1-l_{\Delta P1}) \\ P_{L\Delta P2}/(1-l_{\Delta P2}) \\ P_{L\Delta P3}/(1-\varepsilon l_{\Delta P3}) \end{pmatrix} \quad (1 > l_{\Delta P1}), \qquad \mathbf{s}_{L\Delta D} = \begin{pmatrix} 1 \\ 0 \\ 0 \\ 0 \end{pmatrix}, \quad (13)$$

both with the same eigenvalue $m_{00}$.

Moreover, since $\mathbf{M}_{L\Delta P}$ and $\mathbf{M}_{L\Delta D}$ are Mueller matrices, $\mathbf{M}_{L\Delta P}^T$ and $\mathbf{M}_{L\Delta D}^T$ are also necessarily Mueller matrices, with respective eigenstates

$$\mathbf{t}_{L\Delta P} = \begin{pmatrix} 1 \\ D_{L\Delta D1}/(1-l_{\Delta D1}) \\ D_{L\Delta D2}/(1-l_{\Delta D2}) \\ D_{L\Delta D3}/(1-\varepsilon l_{\Delta D3}) \end{pmatrix} \quad (1 > l_{\Delta D1}), \qquad \mathbf{t}_{L\Delta D} = \begin{pmatrix} 1 \\ 0 \\ 0 \\ 0 \end{pmatrix}. \quad (14)$$

Note that the physical realizability of **M** entails the conditions $l_{\Delta P1} \leq 1$ and $l_{\Delta D1} \leq 1$, otherwise the states $\mathbf{s}_{L\Delta P}$ and $\mathbf{t}_{L\Delta P}$ become unphysical. Furthermore, when $l_{\Delta P1} = 1$, then $P_{L\Delta P1} = 0$, and analogously, $l_{\Delta D1} = 1$ implies $D_{L\Delta D1} = 0$.

Since $\mathbf{s}_{L\Delta P}$ and $\mathbf{t}_{L\Delta P}$ are Stokes vectors, the following inequalities always hold

$$\frac{P_{L\Delta P1}^2}{(1-l_{\Delta P1})^2} + \frac{P_{L\Delta P2}^2}{(1-l_{\Delta P2})^2} + \frac{P_{L\Delta P3}^2}{(1-\varepsilon l_{\Delta P3})^2} \leq 1$$
$$\frac{D_{L\Delta D1}^2}{(1-l_{\Delta D1})^2} + \frac{D_{L\Delta D2}^2}{(1-l_{\Delta D2})^2} + \frac{D_{L\Delta D3}^2}{(1-\varepsilon l_{\Delta D3})^2} \leq 1$$
(15)

## 4 Characterization of a Mueller matrix through its reduced forms#

Let us first recall that the concept of physical Mueller matrix **M** relies on the fact that it can always be considered as a convex sum of nondepolarizing Mueller matrices [13]. Moreover, **M** has a biunivocal relation with its associated coherency matrix **C** defined as

$$\mathbf{C}(\mathbf{M}) = \frac{1}{4}\begin{pmatrix} m_{00}+m_{11}\\ +m_{22}+m_{33} & m_{01}+m_{10}\\ -i(m_{23}-m_{32}) & m_{02}+m_{20}\\ +i(m_{13}-m_{31}) & m_{03}+m_{30}\\ -i(m_{12}-m_{21}) \\ m_{01}+m_{10}\\ +i(m_{23}-m_{32}) & m_{00}+m_{11}\\ -m_{22}-m_{33} & m_{12}+m_{21}\\ +i(m_{03}-m_{30}) & m_{13}+m_{31}\\ -i(m_{02}-m_{20}) \\ m_{02}+m_{20}\\ -i(m_{13}-m_{31}) & m_{12}+m_{21}\\ -i(m_{03}-m_{30}) & m_{00}-m_{11}\\ +m_{22}-m_{33} & m_{23}+m_{32}\\ +i(m_{01}-m_{10}) \\ m_{03}+m_{30}\\ +i(m_{12}-m_{21}) & m_{13}+m_{31}\\ +i(m_{02}-m_{20}) & m_{23}+m_{32}\\ -i(m_{01}-m_{10}) & m_{00}-m_{11}\\ -m_{22}+m_{33} \end{pmatrix}.$$
(16)

Due to its statistical nature, **C** is a positive semidefinite Hermitian matrix [14]. That is, the eigenvalues of **C** are necessarily nonnegative or, equivalently, the leading principal minors of **C** are nonnegative (*covariance conditions*).

Furthermore, it is also well known that *dual retarder transformations* of the form $\mathbf{M}_{R1} = \mathbf{M}_{R2}\mathbf{M}\mathbf{M}_{R1}$, where $\mathbf{M}_{R1}$ and $\mathbf{M}_{R2}$ are orthogonal Mueller matrices (i.e., corresponding to respective retarders) do not affect the covariance conditions, and thus the transformed matrix $\mathbf{M}'$ is a Mueller matrix inasmuch **M** is [5,6,15].

Let us now observe that the transformation (8), can be arranged in the following manner,

$$\mathbf{M} = m_{00}\,\mathbf{M}_{Rf}\,\hat{\mathbf{M}}_{L\Delta P}\,\mathbf{M}_{Rf}^T\,\mathbf{M}_R\,\hat{\mathbf{M}}_D = m_{00}\,\mathbf{M}_{Rf}\left(\hat{\mathbf{M}}_{L\Delta P}\,\hat{\mathbf{M}}_D'\right)\mathbf{M}_R',$$
$$\left(\mathbf{M}_R' \equiv \mathbf{M}_{Rf}^T\,\mathbf{M}_R \quad \hat{\mathbf{M}}_D' = \mathbf{M}_R'\,\hat{\mathbf{M}}_D\,\mathbf{M}_R'^T\right).$$
(17)    (7.17)

This decomposition has the form of a dual retarder transformation of the Mueller matrix $\left(\hat{\mathbf{M}}_{L\Delta P}\,\hat{\mathbf{M}}_D'\right)$. As seen previously, all serial components of decomposition (17) can be calculated from **M**, and consequently, it can be stated that **M** satisfies the covariance conditions if, and only if, $\left(\hat{\mathbf{M}}_{L\Delta P}\,\hat{\mathbf{M}}_D'\right)$ satisfies them. In summary, a normalized $4 \times 4$ real matrix $\hat{\mathbf{M}}$ satisfies the covariance conditions if and only if it can be reduced through the dual retarder transformation (17) into a matrix of the form $\hat{\mathbf{M}}_{L\Delta P}\hat{\mathbf{M}}_D'$, where $\hat{\mathbf{M}}_D'$ is the normalized Mueller matrix of a diattenuator, and the leading principal minors of the coherency matrix $\mathbf{C}_{L\Delta P}$ associated with $\hat{\mathbf{M}}_{L\Delta P}$, are nonnegative.

The explicit expression of $\mathbf{C}_{L\Delta P}$ is

$$\mathbf{C}_{L\Delta P} = \frac{1}{4} \begin{pmatrix} 4c_{L\Delta P0} & P_{L\Delta P1} & P_{L\Delta P2} & P_{L\Delta P3} \\ P_{L\Delta P1} & 4c_{L\Delta P1} & -iP_{L\Delta P3} & iP_{L\Delta P2} \\ P_{L\Delta P2} & iP_{L\Delta P3} & 4c_{L\Delta P2} & -iP_{L\Delta P1} \\ P_{L\Delta P3} & -iP_{L\Delta P2} & iP_{L\Delta P1} & 4c_{L\Delta P3} \end{pmatrix} \quad (18)$$

$$\begin{pmatrix} 4c_{L\Delta P0} \equiv 1 + l_{\Delta P1} + l_{\Delta P2} + \varepsilon l_{\Delta P3} & 4c_{L\Delta P1} \equiv 1 + l_{\Delta P1} - l_{\Delta P2} - \varepsilon l_{\Delta P3} \\ 4c_{L\Delta P2} \equiv 1 - l_{\Delta P1} + l_{\Delta P2} - \varepsilon l_{\Delta P3} & 4c_{L\Delta P3} \equiv 1 - l_{\Delta P1} - l_{\Delta P2} + \varepsilon l_{\Delta P3} \end{pmatrix}$$

while the explicit expressions for the nonnegativity of the principal minors of $\mathbf{C}_{L\Delta P}$ are the following

$$16 c_{L\Delta P0} c_{L\Delta P1} \geq P_{L\Delta P1}^2$$

$$16 c_{L\Delta P0} c_{L\Delta P1} c_{L\Delta P2} \geq c_{L\Delta P2} P_{L\Delta P1}^2 + c_{L\Delta P1} P_{A\Delta P2}^2 + c_{L\Delta P3} P_{A\Delta P3}^2$$

$$16 c_{L\Delta P0} c_{L\Delta P1} c_{L\Delta P2} c_{L\Delta P3} - \frac{1}{16} P_{L\Delta P}^2 \quad (19)$$

$$\geq \left( c_{L\Delta P0} c_{L\Delta P1} + c_{L\Delta P2} c_{L\Delta P3} \right) P_{L\Delta P1}^2 + \left( c_{L\Delta P0} c_{L\Delta P2} + c_{L\Delta P1} c_{L\Delta P3} \right) P_{L\Delta P2}^2 + \left( c_{L\Delta P0} c_{L\Delta P3} + c_{L\Delta P1} c_{L\Delta P2} \right) P_{L\Delta P3}^2$$

Therefore, the reduced form $\hat{\mathbf{M}}_{L\Delta P}$ of $\mathbf{M}$ provides particularly simple analytical expressions for the covariance conditions of Mueller matrices. The difference with respect to the characterization obtained in Ref. [15] based on the arrow form $\mathbf{M}_A$ of $\mathbf{M}$, is that the above characterization involves additionally the normal diattenuator $\hat{\mathbf{M}}'_D$, whose particular characterization is, in turn, very simple and well known.

Obviously, a completely analogous procedure for the characterization of Mueller matrices can be performed starting from the reverse reduced form $\hat{\mathbf{M}}_{L\Delta D}$ of $\mathbf{M}$ instead of from $\hat{\mathbf{M}}_{L\Delta P}$.

**5 Singular Mueller matrices**

This section is devoted to a brief analysis of the different families of singular Mueller matrices, for whose classification the reduced forms are particularly useful.

From Eqs. (8) and (10) the determinant of $\mathbf{M}$ can be expressed in the following forms

$$\det \mathbf{M} = m_{00}^4 \det \hat{\mathbf{M}}_{L\Delta P} \det \hat{\mathbf{M}}_D = \varepsilon m_{00}^4 l_{\Delta P1} l_{\Delta P2} l_{\Delta P3} \left( 1 - D^2 \right)$$

$$\det \mathbf{M} = m_{00}^4 \det \hat{\mathbf{M}}_P \det \hat{\mathbf{M}}_{L\Delta D} = \varepsilon m_{00}^4 l_{\Delta D1} l_{\Delta D2} l_{\Delta D3} \left( 1 - P^2 \right) \quad (20)$$

and consequently, leaving aside the trivial case of the zero Mueller matrix (i.e., $m_{00} = 0$), $\mathbf{M}$ satisfies $\det \mathbf{M} = 0$ if and only if at least one of the following conditions holds: $P = 1$ (i.e., $\mathbf{M}$ corresponds to a *depolarizing polarizer* [1]); $D = 1$ ($\mathbf{M}$ corresponds to a *depolarizing analyzer* [1]), or $l_{\Delta P3} = l_{\Delta D3} = 0$ ($\mathbf{M}$ corresponds to a *singular depolarizer* [16]).

The last case holds for media for which the fact that its associated Mueller matrix is singular is caused by the depolarization properties rather than including a polarizer (with $P = D = 1$) as a serial component. Singular depolarizers are studied in a separate paper [16].

**Conclusion**

Two dual reduced forms of a generic Mueller matrix have been defined that provide an additional way for the study and interpretation of some interesting properties of Mueller matrices. In particular, the power of these representations has been proved for characterizing the covariance properties of Mueller matrices as well as for the identification of the causes for which a Mueller matrix is singular.

**Acknowledgement**

This research was supported by Ministerio de Economía y Competitividad and the European Union, grants FIS2011-22496 and FIS2014-58303-P, and by Gobierno de Aragón, group E99.